\documentclass{rspublic}
\usepackage{amsmath}
\usepackage{amsfonts}
\usepackage{amssymb}
\usepackage[Symbol]{upgreek}
\usepackage[nointegrals]{wasysym}
\usepackage{mathrsfs}
\usepackage{graphicx}
\newcommand{\real}{{\mathbb R}}
\newcommand{\bfu}{{\bf u}}
\newcommand{\bfv}{{\bf v}}

\newcommand{\bfx}{{\bf x}}
\newcommand{\bfy}{{\bf y}}
\newcommand{\bfk}{{\bf k}}
\newcommand{\bfL}{{\bf L}}
\newcommand{\hfu}{{\bf \hat{u}}}
\newcommand{\hu}{{\hat{u}}}
\newcommand{\Upo}{{\Omega}}
\newcommand{\bdy}{{\partial \Omega}}

\begin{document}

\title[Periodic Navier-Stokes]{\large The Navier-Stokes equations in periodic domains}

\author[F. Lam]{F. Lam}

\affiliation{ }

\label{firstpage}

\maketitle

\begin{abstract}{Navier-Stokes; Viscosity; Turbulence; Finite Energy}
In the present technical note, we establish that the setting of the primitive variables of the unsteady incompressible fluid dynamics is ill-formulated in spatially periodic domains as the specification of the boundary velocity is too broad to sidestep time-dependency and approximation errors. As an illustration, we show that the Taylor-Green solution in planes suffers from the Hadamard-divergence, and the ABC flow in cubes is non-unique. In direct numerical simulations of homogeneous turbulence with no corrective precautions on the boundary values, our assertion helps us understand the well-experienced nuisances, such as slow rates of convergence in energy dissipation, fluctuations in the statistics moments, or spontaneous surges in the time-averaged flow quantities. In particular, vorticity dynamics is not described by singular integral equations. 
\end{abstract}

\section{Background}

In incompressible real fluids, the Navier-Stokes equations of motion are derived from the principles of momentum conservation, and mass  
\begin{equation} \label{ns}
	\partial_t \bfu - \nu \Delta \bfu = - (\bfu \cdot \nabla) \bfu  - {\rho}^{-1} \nabla p,\;\;\; \nabla{\cdot}\bfu=0,
\end{equation}
where $\Delta$ stands for the Laplacian, the vector quantity $\bfu(\bfx,t)=(u,v,w)(\bfx,t)$ is the velocity ($\bfx=(x,y,z)$), the scalar $p(\bfx,t)$ the pressure. We also use the tensor notation, $\bfu(\bfx,t)=(u_i(\bfx,t)),\bfx=(x_i),i=1,2,3$.
The symbol $\nu(=\mu/\rho)$ denotes the kinematic viscosity, where $\rho$ and $\mu$ are the density and the viscosity of the fluid respectively. For inviscid flows $\nu=0$, the system is the Euler equations. Taking divergence of (\ref{ns}) and using the continuity, we obtain a Poisson's equation for the pressure 
\begin{equation} \label{ppoi}
	\Delta p(\bfx) = - \rho \nabla\cdot \big( (\bfu{\cdot}\nabla) \bfu \big)(\bfx), 
\end{equation}
which holds at every instant of time $t$. The initial condition for the velocity is solenoidal and smooth
\begin{equation} \label{ic}
 \bfu(\bfx,0)=\bfu_0(\bfx)\; (\in C^{\infty}(\Upo)).
\end{equation} 

We are interested in finite-energy initial value problems subject to the initial data (\ref{ic}). To focus on the key issues of our note, let us consider viscous flows in a periodic cube ($\Upo'$) with period $\bfL=(L_x,\;L_y,\;L_z)$, $0 \leq x,y,z \leq \bfL$. The boundary conditions take the form of
\begin{equation} \label{pbc}
\bfu(\bfx)\;=\;\bfu(\bfx+\bfL).
\end{equation}
By default, a cube is defined by $L=L_x=L_y=L_z$.
Here the actual magnitudes in the velocity are {\it not} fixed; they are allowed to vary over the flow evolution in general. As the pressure only plays an auxiliary role in fluid motion, there must be no boundary conditions on the pressure. Otherwise, the dynamics is over-specified since the pressure can be eliminated by the incompressibility constraint, reducing (\ref{ns}) to a consistent system of three equations in three unknown velocities.

\section{Taylor-Green vortices}

Consider an example of viscous flow in a square box ($0\leq x,y \leq 2 \pi$) with boundary data 
\begin{equation} \label{tgpbc}
\begin{split}
	u(x,0)&=u(x,2 \pi)=0, \;\;\; u(0,y)=u(2 \pi,y),\\
	v(0,y)&=v(2 \pi,y)=0,\;\;\; v(x,0)=v(x,2 \pi).
	\end{split}
\end{equation}
It has been known for long that the Taylor-Green eddy, 
\begin{equation} \label{tg}
\begin{split}
	u(\bfx,t)&=\cos(x) \sin(y) f(t),\\
		v(\bfx,t)&=-\sin(x)\cos(y) f(t),\\
		p/\rho(\bfx,t)&=-\big(\cos(2x)+\cos(2y)\:\big)f^2(t)/4,
	\end{split}
\end{equation}
where $f=\exp(-2\nu t)$, is an exact solution of the planar equations (\ref{ns}). Given the velocity $(u,v)$, the vorticity and the stream function are found to be
\begin{equation*} 
 \zeta=-2\cos(2x) \cos(2y) f(t),\;\;\mbox{and}, \;\; \psi=-\cos(x) \cos(y) f(t)
\end{equation*}
respectively.

Now we demonstrate the existence of a generalised vortex structure.
First, we select the velocity and the pressure,
\begin{equation} \label{sp1}
\begin{split}
\tilde{u}(\bfx,t)&=\cos(m x) \sin(m y) f(t),\\
		\tilde{v}(\bfx,t)&=-\sin(m x)\cos(m y) f(t),\\
		\tilde{p}/\rho(\bfx,t)&=-\big(\cos(2mx)+\cos(2my)\:\big)f^2(t)/4,
	\end{split}
\end{equation}
where the time-dependent factor $f(t)=\exp(- 2 m^2 \nu t)$, and $m$ is an arbitrary integer. Note that the periodicity of $(u,v)$ is preserved, and the homogeneous boundary data are respected for $t>0$. When the viscosity is low, these solutions are mild oscillations at moderate values of $m$. Second, we observe that the exponential decay has other choices as long as $f(t)$ satisfies
\begin{equation} \label{tt}
	\rd f/\rd t + 2 m^2 \nu f = 0.
\end{equation}
An interesting case is $m \rightarrow m/\sqrt{\nu}$ (assumed an integer) so that the decay is independent of the viscosity. In particular, the solutions,
\begin{equation}  \label{sp2}
\begin{split}
	\tilde{u}(\bfx,t) &= \cos(m x/\sqrt{\nu}) \sin(m y/\sqrt{\nu}) \exp(-2 m^2 t),\\
	\tilde{v}(\bfx,t) & = -\sin(m x/\sqrt{\nu}) \cos(m y/\sqrt{\nu}) \exp(-2 m^2 t),
		\end{split}
\end{equation}
define a velocity field which may be rapidly oscillating with slow decay even for small values of $m$, soon after the start at locations $x\approx0$ or $y\approx 2\pi$ (say). At least for some viscosity, these high-frequency fluctuations must constitute poor representations for any genuine flow structure because they are the ill-defined solutions in the sense of Hadamard (1964).

In three space dimensions, the initial Taylor-Green velocity is given by
\begin{equation} \label{tg3}
\begin{split}
	\bfu_0(\bfx)=\big( \: A \cos(a x) \sin(b y) \sin(cz), \; B\sin(ax) & \cos(by) \sin(cz), \\
	\quad & C\sin(ax)\sin(by) \cos(cz) \: \big),
	\end{split}
\end{equation}
where the incompressibility demands $Aa+Bb+Cc=0$. Closed-form algebraic expressions of solution $(\bfu,p)(\bfx,t)$ are not known to exist (cf. Taylor \& Green 1937; Goldstein 1940). On the other hand, nearly all the computations by numerical approximations are undertaken subject to the periodic boundary condition (\ref{pbc}) so that the subsequent solutions enjoy a high degree of symmetry and accuracy, as advocated by practitioners of spectral methods. In contrast to the planar case, a key consequence is that the initial homogeneous data at the edges of the planes normal to each direction are {\it not} preserved over time. At any $t>0$, the boundary velocities can be {\it de facto} functions of the spectral modes, viscosity, as well as numerical error amass. It is fair to assert that the dynamic evolution must be more vulnerable to the Hadamard-divergence, as characterised by the sine-cosine oscillations with relaxed constraints on the boundaries (cf. the scenarios implied in (\ref{sp2})). Does the initial-boundary value problem remain well-posed?

\section{General spurts}

Our discussion in the preceding section inspires space-time-dependent transforms on the three-dimensional equations. In periodic domains with periodic boundary conditions, a `spurt' transform, $(\tilde{\bfu}, \tilde{p})$, is a set of special Navier-Stokes functions such that, if $(\bfu,p)$ satisfies the Navier-Stokes dynamics, so does $(\bfu+\tilde{\bfu}, p+\tilde{p})$. Note that there are no data for pressure and for vorticity on $\bdy'$ (unless we force them). First, the energy is uncertain,
\begin{equation} \label{e2}
\frac{1}{2}\frac{\rd}{\rd t}\int_{\Upo'}  |\bfu|^2 \; \rd \bfx + \nu \int_{\Upo'} |\nabla \bfu|^2 \; \rd \bfx = [\bfu p]_{\bdy'} + \nu [\bfu \nabla \bfu]_{\bdy'} \neq 0,
\end{equation}
regardless of flow regularity, unless $\bfu$ vanishes on the boundary. However, specification of periodic data is designed to allow for non-zero data. 

\subsection*{Time-wise spurt}

A spurt perturbation, $(\tilde{\bfu}, \tilde{p})$, is a set of specific functions that satisfy the dynamic equation, (\ref{ns}), and the periodic boundary conditions. The following spurt:
\begin{equation} \label{spt}
\begin{split}
\tilde{u}(\bfx,t) & = C_1 \: f(t),\;\;\; \tilde{v}(\bfx,t) = C_2 \: g(t),\;\;\; \tilde{w}(\bfx,t) = C_3 \: h(t),\\
	\tilde{p}(\bfx,t)/\rho & = - C_1 \: xf' - C_2 \: yg' - C_3 \: zh',
	\end{split}
\end{equation}
is a solution for viscosity $\mu \geq 0$. Constants, $C_1,C_2$ and $C_3$, are arbitrary and finite. The spurt functions, $f,g,h$, are assumed to be smooth and bounded, and
\begin{equation} \label{sqt}
f(0)=g(0)=h(0)=0,\;\;\;f'(0)=g'(0)=h'(0)=0.
\end{equation}
They can be arbitrarily chosen, as long as the energy is bounded.

Let $(\bfu,p)$ be primitive solutions with initial data $\bfu(\bfx,0)$. Then the following pair, 
\begin{equation}
(\;u(\bfx,t) + \tilde{u}, \; v(\bfx,t) + \tilde{v},\; w(\bfx,t) + \tilde{w}, \;\;p(\bfx,t)+ \tilde{p}\;),
\end{equation}
satisfy the equations of motion, provided the momentum defines the pressure
\begin{equation} \label{pdn}
\partial_t \tilde{\bfu}+(\tilde{\bfu}{\cdot}\nabla)\bfu = -\nabla \tilde{p}/\rho,
\end{equation}
where $\tilde{\bfu}=\tilde{\bfu}_0$, as given by (\ref{spt}). The constraints (\ref{sqt}) preserve initial data $\bfu(\bfx,0)$, as well as starting pressure. By the continuity, the spurt pressure is determined by Laplace's equation,
$\Delta \tilde{p} =0$, plus six boundary conditions given by (\ref{pdn}). Hence $\nabla\tilde{p}$ is unique by solving the Neumann problem using the method of Green's function. We assume that solutions ($\bfu,p$) ensure compatible boundary data. Explicitly, we require
\begin{align*}
\;u_x(0,y,z)&=u_x(L_x,y,z),& u_y(0,y,z)&=u_y(L_x,y,z),& u_z(0,y,z)&=u_z(L_x,y,z),\\
\;v_x(x,0,z)&=v_x(x,L_y,z),& v_y(x,0,z)&=v_y(x,L_y,z),& v_z(x,0,z)&=v_z(x,L_y,z),\\
\;w_x(x,y,0)&=w_x(x,y,L_z),& w_y(x,y,0)&=w_y(0,y,L_z),& w_z(x,y,0)&=w_z(x,y,L_z),
\end{align*}
for admissible $\nabla \tilde{p}|_{\bdy'}$.

Although it is difficult to present general periodic solutions, we gain considerable insight into the nature of periodic solutions using a general Taylor-Green vortex flow as illustration. Let $(\bfu,p)$ be given by $\bfu(\bfx,t)=\tilde{\bfu}(\bfx)\exp(- \nu t \kappa^2)$, where
\begin{equation} \label{vtg}
\begin{split}
\tilde{u}(\bfx,t)&= A \sin (2 \pi \alpha x) \cos (2 \pi \beta y) \cos (2 \pi \gamma z),\\
\tilde{v}(\bfx,t)&= B \cos (2 \pi \alpha x) \sin (2 \pi \beta y) \cos (2 \pi \gamma z),\\
\tilde{w}(\bfx,t)&= C \cos (2 \pi \alpha x) \cos (2 \pi \beta y) \sin (2 \pi \gamma z),\\
\end{split}
\end{equation}
$(\alpha,\beta,\gamma)=(l/L_x, m/L_y, n/L_z)$, for any integers, $l,m,n = 0,\pm1,\pm2,\pm3,\cdots$, and $\kappa^2=4\pi^2(\alpha^2+\beta^2+\gamma^2)$. The continuity demands $\alpha A + \beta B + \gamma C=0$. As usual, the pressure is found from, $\Delta p = -\rho \nabla{\cdot} \big((\bfu{\cdot}\nabla)\bfu\big)$. This vortical flow ensures that the boundary conditions for spurt $\nabla\tilde{p}$ are all compatible.

\subsection*{Space-wise spurt}

There are rotational spurts that are incompressible solutions in the domain $\Upo'$. One example is given by
\begin{equation} 
\tilde{u} = \sin(2 \pi \sigma y) \exp(-\nu \sigma^2 t),\;\;\; \tilde{v} =\tilde{w} = \tilde{p} = 0,
\end{equation}
where $\sigma=m/L_y, m=0,\pm1,\pm2,\pm3, \cdots$. 

Consider the superimposed solution ($\bfu+\tilde{\bfu},p+\tilde{p}$), where $\tilde{v}=\tilde{w}=0$, and
\begin{equation} \label{spp}
\tilde{u}=\begin{cases}
\; Q(t_0)\sin(2 \pi \sigma y) \exp(-\nu \sigma^2 t), & t \geq t_0,\\
\; 0,& 0 \leq t < t_0.
\end{cases}
\end{equation}
The equations of motion become $(\tilde{\bfu}{\cdot}\nabla)\bfu + (\bfu{\cdot}\nabla)\tilde{\bfu}= -\nabla \tilde{p}/\rho$. We have to determine the pressure from
\begin{equation}
\Delta \tilde{p} = - 4 \pi \rho \sigma \cos(2 \pi \sigma y) \partial_x v,
\end{equation}
subject to 4 non-zero data, $\partial_x \tilde{p}|_{x=0,L_x}$, and $\partial_z \tilde{p}|_{z=0,L_z}$. There are no specific reasons to limit the initiation and size of ($\tilde{u},0,0,\tilde{p}$). At small viscosity in particular, a sine-wave introduced at $t=t_0$ gives rise to a pulsation in $\bfu$'s transient. 

In numerical solutions by spectral method, functional forms of (\ref{spt}) or (\ref{spp}) emulate discretisation errors in time-marching. Any apparent spiky or intermittent variations in flow configuration are not end results of a singularity. These examples are adequate to demonstrate how unpredictable periodic solutions can be. In contrast, there are no periodic data for vorticity. Because of lack of uniqueness, primitive solutions, $(\bfu,p)$, cannot offer satisfactory explanations to flow physics in periodic domains. Lastly, the spurts are still effective for inviscid flows described by the Euler equations if viscosity is formally taken to be zero.

The 2$d$ weak solutions constructed by Shnirelman (1997) and the generalised weak solutions with decreasing energy (Shnirelman 2000) are believed to intersect with a well-defined time-dependent smooth solution plus arbitrary numbers of the (finite) spurts. Also we have made a thorough check on the `blow-up' yes-list of Gibbon (2008): all the finite-time singularity computations have been performed on periodic boxes with periodic boundary conditions. Moreover, mesh convergence validation on the events prior to the inception of a singularity was almost non-existent in these numerical works, and no efforts have been made to monitor the boundary values over the time-marching. (In those simulations where the periodic pressure is imposed, the dynamics is over-determined and hence mathematically inconsistent.) Similarly, in the method of analytic strip, the Euler equations are analytically continued into complex periodic domain, see, for instance, equations (7)-(9) of Frisch {\it et al}. (2003), if we generalise the primitive variables $(\bfu,p)$ to $\bfu(\bfx,t)=\bfu_r(\bfx,t)+\ri \:\bfu_i(\bfx,t)$, and $p(\bfx,t)=p_r(\bfx,t)+\ri \:p_i(\bfx,t)$, $\ri=\sqrt{-1}$. With periodic boundaries, we can devise, by analogy to equation (\ref{spt}), two distinct spurt quadruplets, ($\tilde{\bfu}_r,\tilde{p}_r$) and ($\tilde{\bfu}_i,\tilde{p}_i$), for the real and imaginary parts. Unavoidably, the analytic-strip approach to Euler flow inherits the spurt-inflicted non-uniqueness in its solutions. 

In parallel, it is commonly observed in laboratory or Nature that, for given initial conditions, incompressible eddies mutually interact, agglomerate, break up, mingle, merge, diffuse, and attenuate, with finite enstrophy and energy. While recognising a dedicated endeavour of a mass-scale direct numerical simulation by Kaneda {\it et al}. (2003), the conclusion on the existence of a non-zero energy dissipation in the limit $\nu \rightarrow 0$ must be considered as indicative, if not incorrect, as a static spurt may have been enmeshed in the discretisation. The viability of (\ref{spt}) extends to the computational theory of projection methods (see, for instance, Chorin 1969) where the convergence analyses must be reappraised, as an indeterminate velocity spurt may be consigned to each component at every iteration, thus rendering the finite-difference scheme unquantifiable. By the same token, the trefoil vortex rings (Kerr 2018) were simulated in a periodic box with free-slip boundaries. The circulation field cannot be immune to the spurts which corrupt any emerging vortices to all length scales (in light of the well-known density theorem), regardless of the box size; therefore, his results of vortex reconnection with local self-similarity, and the related discussion of scaling bounds, are utterly misleading.

\section{Mistimed ABC flow}

A specific type of incompressible flow is discussed in Dombre {\it et al.} (1986) with initial conditions ($L=2 \pi$)
\begin{equation} \label{abc0}
 \bfu_0(\bfx)=\big( \: \sin(mz){+}\cos(my), \: \sin(mx){+}\cos(mz), \: \sin(my){+}\cos(mx)\: \big).
\end{equation}
The initial-boundary value problem can be exactly solved subject to non-zero periodic boundary data.
A straightforward computation shows that the solutions are given by
\begin{equation} \label{abc}
\begin{split}
\bfu(\bfx,t)&=\big( \sin(mz){+}\cos(my), \sin(mx){+}\cos(mz), \sin(my){+}\cos(mx)\big) f(t),\\
p(\bfx,t)/\rho&=-\big( \cos(mx)\sin(my) {+} \sin(mx) \cos(mz) {+} \cos(my) \sin(mz) \big)f^2(t),
	\end{split}
\end{equation}
where $f(t)$ denotes one of the solutions of (\ref{tt}), and $f(t{=}0){=}1$. Evidently, the ABC flows are susceptible to the instability {\it \'a la} Hadamard which is seen only as a possibility. What is less understood is the fact that every flow field of (\ref{abc}) is {\it invariant} to the spurt transform (\ref{spt}) because the addition of the spurt functions preserves the $\bfu$-boundary periodicity. Also the suitably-chosen $(\tilde{\bfu}{-}\tilde{p})$ solutions do not modify the initial data. The essence is that we cannot rely on extensive computations or symmetry arguments to justify ABC's simplicity and diversity without fair appreciation of the spurts. No matter how sophisticated our numerical algorithms appear to be, the numerical outputs largely contain unquantified temporal ingredients of obscurity. Within the parameters of practical interests, the claimed singular, chaotic or unpredictable phenomena exhibited in the ABC flows must have limited meaning, if not irrelevant to the continuum physics. 

\section{Fourier modes}

In periodic domain $\Upo'$, we represent velocity as Fourier series 
\begin{equation} \label{fs}
	\bfu(\bfx,t)=\sum_{\bfk} \;\exp(\;\ri\; \bfk \cdot \bfx) \hfu(\bfk,t),
\end{equation}
where the discrete wave-numbers $\bfk=\big(2\pi l/L_x, 2 \pi m/L_y, 2 \pi n/L_y\big)$,
with integers $l,m,n$ $=0,\pm1,\pm2,\pm3,\cdots$. The coefficients $\hfu$ are the (complex) amplitudes of the modes designated by the wave-number $\bfk$, and $\hfu(\bfk,t)=\hfu^*(-\bfk,t)$, where the superscript asterisk stands for the complex conjugate. The Fourier transform of $\bfu(\bfx,t)$ is defined by
\begin{equation} \label{ftv}
	\hfu(\bfk,t)=\frac{1}{\cal V}\int_{\Upo'}\exp(- \;\ri \;\bfk \cdot \bfx)\bfu(\bfx,t) \;\rd \bfx,
\end{equation}
where the volume of the box ${\cal V}=L_xL_yL_z$.
We omit the analogous expression for pressure. In wave-number space, the continuity becomes
\begin{equation} \label{wct}
	k_l\hu_l(\bfk,t)=0.
\end{equation}
The Navier-Stokes momentum is transformed into
\begin{equation} \label{wns}
	\bigg( \frac{\rd}{\rd t}+ \nu k^2 \bigg)\hu_l(\bfk,t)+{\cal V}\;\ri k_n\bigg(\delta_{lm}{-}\frac{k_lk_m}{k^2}\bigg) \sum_{\bfk'+\bfk''=\bfk}\hu_m(\bfk',t)\:\hu_n(\bfk'',t)=0,
\end{equation}
where $\delta_{ij}$ stands for Kronecker's delta. Now, the non-linearity is characterised by the interactions among the modes, $\bfk$, $\bfk'$ and $\bfk''$. A conceptual prerequisite is assumed: wave-numbers become the surrogates of the physical length scales.

There is a hindrance to the Fourier approach. Let us configure flow-field in
\begin{equation} \label{cfv}
(\bfu+\bfv,p+q)(\bfx,t),
\end{equation}
where $\bfv=\tilde{\bfu}_0$ as given by (\ref{spt}). The principle of momentum conservation now reads
\begin{equation} \label{nspt}
\partial_t (\bfu + \bfv) - \nu \Delta \bfu + (\bfu {\cdot} \nabla) \bfu + {\rho}^{-1} \nabla (p+q) +  (\bfv{\cdot}\nabla)\bfu=0.
\end{equation}
Abbreviate the transformed equation (\ref{wns}) by ${\cal F}(NS)=0$. We carry out the Fourier transform on (\ref{nspt}). The result is summarised in
\begin{equation} \label{wnsp}
{\cal F}(NS) = - \ri k_l \:(\hat{q}+ v_l \hu_l) = 0,
\end{equation} 
as long as pressure $\hat{q}$ is fixed for known $\hfu$. Consequently, no solution pair constructed by (\ref{fs}) in terms of the Fourier components of (\ref{wct})-(\ref{wns}), agree with velocity field (\ref{cfv}) which is arbitrary. This conclusion holds for any viscosity $\mu \geq 0$.

\section{Dissipation anomaly}

Anomalous dissipation (Onsager, 1949): At the limit $\nu \rightarrow 0$, the solutions of the Navier-Stokes, or effectively Euler's equations, in the periodic setting are distributions, and the kinetic energy remains dissipated in nearly inviscid turbulence, because of the non-smoothness of the velocity field, hence violating the law of energy conservation.

In periodic domains, where turbulence is regarded as isotropic or homogeneous, the energy is given by
\begin{equation*}
\frac{1}{2} \frac{\rd}{\rd t} \int_{\Upo'} |\bfu|^2\; \rd \bfx = - \nu \int_{\Upo'} |\upomega|^2\; \rd \bfx, 
\end{equation*}
provided that conditions, $\bfu|_{\bdy'}=0$ or $p|_{\bdy'}$ and $\nabla\bfu|_{\bdy'}$, are forced. Moreover, the above integration is meaningful only when the velocity has certain regularity, for instance, $\bfu, \upomega \in C^1(\Upo') \cap L^2(\Upo')$. An assumption was made (p.286 of Onsager, 1949) that, in the limit $\nu \rightarrow 0$, the derivation of energy conservation was no longer applicable for non-differentiable velocity, because the velocity correlation in physical space could be H{\"o}lder continuous,
\begin{equation} \label{hc}
|\bfu(\bfx+\bfx')-\bfu(\bfx)| < C |\bfx'|^{\lambda},\;\;\; \lambda < 1/3,
\end{equation}
where Onsager merely contemplated the problem of regularity\textemdash but neglected the under-determinacy due to the periodic setting. He argued that there exists a cascade of energy among wave interactions of $\pm\bfk_l$ and $\pm\bfk_m$. The cascade is made of consecutive stages, instigated by velocity gradients or shears. Over each stage, the wave numbers would increase two-fold, giving rise to repeated binary fissions over the process of energy transfer from low to high wave-numbers. In the absence of regularity, there are no upper bounds on the size of wave numbers as $\bfu \in L^1_{loc}$. With increasing wave numbers, the velocities become so small and discontinuous, that the fluid elements themselves start losing energy over continual mutual bombardments, as if there is no viscosity. 

This idea appears to be a reductionist view which grossly simplifies the role of the non-linearity. As shown before, the velocity is ill-posed in the periodic formulation in the primitive variables, that downplay the rate of shear (as measured by palinstrophy, $\|\nabla \upomega\|_{L^2}$). It is the periodicity hypothesis which leads to the notions of an inertial range (energy transfer) and dissipative scales (viscous damping) in wave-number space. Their counter-parts in real flow do not exist, see the equivalence of (\ref{wns}) and (\ref{wnsp}). The anticipated wave numbers do not correspond to the physical scales. Since there are no vorticity boundary conditions, evaluating vorticity dynamics in periodic domains must be a misleading alternative. Recall that the rate of averaged energy dissipation, $\rd \bar{u}^2/\rd t \propto \bar{u}^2 \;\bar{u}/L$, is an empirical formula made on dimensional analysis. This estimation may offer explanations for specific experiments but is unsuitable for elucidating dissipation by viscosity whose effectiveness depends on the non-linear production of flow scales. In fact, function (\ref{hc}) is absurd in periodic domains, as the velocity field is non-unique in space or in time.

\section{Remarks}

The significance of the spurts is that the incompressible fluid dynamics in periodic domains with periodic boundary conditions is {\it not well-determined} in the formulation of the primitive pair $(\bfu,p)$. Any theoretical treatment of the dynamics necessitates further justifications on the nature of the boundary values. The present author has been unaware of any definitive work on this apparently trivial but critical matter in the technical literature. It has been a (mis)belief that, on the basis of large-scale numerical computations, the Navier-Stokes dynamics blows up in finite time for suitable initial data of finite energy. Some functional analyses do result in diverged solutions which are often classified into the categories of `weak' non-uniqueness or `wild' solutions.\footnote{If we relax the smoothness requirement on the spurts, we can actually construct a singular Navier-Stokes solution consisting of the Cantor function or the Devil's staircase over time $[0,1]$ (say), because every stair height (normalised) preserves the boundary periodicity, as long as we choose the spurt pressure to be zero almost everywhere.}  Evidently, any analysis without mathematical devices to filter out the spurts cannot be regarded as complete. Our exposition has implications in the numerical simulations of homogeneous turbulence in the box with period $L$. In practice, the spurt functions may be understood as the representation for the numerical errors arising from truncation, spatial discretisation, mesh resolution, finite-precision arithmetic or aliasing procedures. Because the boundary velocities are loosely fixed in (\ref{pbc}), the numerics may have appeared successfully in calculations from time $t_n$ to the next step $t_{n+1}$. The problem is that the errors may initiate and propagate in non-transparent manners. Then the converged `solutions' would contain unquantified jumps in the boundary values due to accumulations. If the numerical surges are frequent and repeat over time, one is tempted to view fluid motions as topologies of multi-fractals which contradict the underlying principles of the Navier-Stokes dynamics. 

To put potentially awkward spurts into perspective, let us consider the examples
\begin{equation} \label{ex}
	g(t)=t^2\: \exp\big(\:t^3/\nu\:\big)\;\; \mbox{or}\;\; h(t)=L\exp\big(\:t^{1+\epsilon} \cos(m/\sqrt{\nu})\:\big)-L, \;\;(\epsilon>0).
\end{equation}
In actual computations, we may not have no full knowledge of the way the spurt functions creep into our numerical approximations, small-viscosity motions starting from (\ref{ic}) subject to (\ref{pbc}) may well run amok over a tiny instant $t>0^+$. In other words, the existence of the spurts may well render a solution ($\bfu,p$) into the class of Hadamard-divergence. Indeed, a solution containing a fraction of $h(t)$ of (\ref{ex}) must appear as incongruous and confusing, possibly giving rise to the familiar characters of a temporal intermittency. 

It has been reported that, in direct numerical simulations of homogeneous turbulence in periodic boxes, certain higher-order moments of enstrophy and energy dissipation converge slowly, if at all. The poor rates of convergence are likely the numerical pathologies related to the presence of mild spurt functions trapped in various spectral modes. It is also known, that these global time-averaged quantities can fluctuate by several orders of magnitude over a particular period of time. One plausible explanation is that the spurt arbitrariness has been out-of-control in the numerics (cf. $g(t)$ in example (\ref{ex})).

In the vorticity formulation, $\upomega(\bfx) = \nabla {\times} \bfu(\bfx)$, the scalar pressure is eliminated from the system (\ref{ns}) so that a compensator for  potential velocity arbitrariness has been removed, as expressed in the conservation law of angular momentum
\begin{equation*} 
	\partial_t \upomega - \nu \Delta \upomega = (\upomega \cdot \nabla) \bfu  - (\bfu \cdot \nabla )\upomega.
\end{equation*}
However, the velocity must be recovered from the di-vorticity $\Delta \bfu = - \nabla{\times} \upomega$. In this scheme, the boundary $\bfu$ must be prescribed and fixed over time. In periodic domains, this specification is tantamount to a loss of periodicity as, now, the problem is to find solutions in bounded domains with Dirichlet data.\footnote{In general, there exist no boundary conditions for vorticity in the Navier-Stokes dynamics, see a review by Gresho (1991). The numerical computations of Ayala, Doering \& Simon (2018) employed (artificial) periodic boundary conditions for the vorticity. Following this (erratic) approach, we may specify vorticity spurts in the periodic setting, for example, $\tilde{\omega}=f(Re_0)$, where function $f$ is non-zero arbitrary but bounded. Alternatively, we may assign $\tilde{\omega}=f(\varepsilon_d)$, where $\varepsilon_d$ denotes the maximum size of the discretisation errors in each mesh resolution. The existence of these spurts implies that $\Delta \tilde{\bfu}=0$, which generates one or more spurt velocity fields at every instant $t$ in the numerics. There are many harmonic functions available for constructing solenoidal velocity spurts in the square domain $L{\times}L$. We notice that no consistent convergence was demonstrated in their simulations. In numerous practical works at low $\nu \sim O(10^{-5} {\sim} 10^{-4})$ with proper velocity boundary data (for example, the no-slip on $C^2$ boundaries), it is well-documented that the palinstrophy in a flow may surge to an order of $O(10^8 {\sim} 10^9)$ in the early phase of evolution. The numerical size of $|\nabla \omega|^2$ is not a matter of concern, as long as the flow remains incompressible; the key criterion is whether the energy, enstrophy as well as palinstrophy, have converged or not in the light of systematically refined mesh-resolutions. Briefly, their results are rather crude and tentative. Above all, their ideas for finite-time singularities directly contradict the well-established global regularity for two-dimensional incompressible viscous flows, see Leray (1934); and Ladyzhenskaya (1959).} 

By the stream function vector $\bfu=\nabla{\times}\uppsi$, where $\Delta\uppsi-\nabla(\nabla{\cdot}\uppsi)=-\upomega$, $\bfu$ can be recovered by solving
\begin{equation} \label{vpo}
\Delta \uppsi = - \upomega,\;\;\; \nabla.\uppsi=0,\;\;\;\bfx \in \real^3,
\end{equation}
assuming $\uppsi$ decays. The attenuation is justified for flows of finite energy. For unique (vector) stream function $\uppsi$, we put $\uppsi=\uppsi'+\nabla\phi'+{\bf c}$, where $\Delta \phi'=-\nabla{\cdot}(\uppsi'+{\bf c})$, and $\bf c$ is any irrotational constant. Conversely, the vector potential is given by 
\begin{equation} \label{vp}
\uppsi(\bfx) = \frac{1}{4 \pi}\int_{\real^3} \frac{1}{|\bfx-\bfy|}\;\upomega(\bfy)\; \rd \bfy=\frac{1}{4 \pi}\int_{\real^3} \frac{1}{\;|\bfy|\;}\;\upomega(\bfx{+}\bfy)\; \rd \bfy.
\end{equation}
The Biot-Savart law shows that gauge choice, $\nabla{\cdot}\uppsi=0$, holds everywhere, provided vorticity is solenoidal, $\nabla{\cdot}\upomega=0$. In view of (\ref{vpo}) and (\ref{vp}), the gauge choice, $\nabla{\cdot}\uppsi=0$, implies $\nabla{\cdot}\upomega=0$. It is evident that vector identity, $\nabla{\times}\nabla{\times}\uppsi=-\upomega$, is inherently solenoidal by formula $\nabla{\cdot}\nabla{\times}{\bf A}=0$. At least, $\partial_x \xi$, $\partial_y \eta$ and $\partial_z \zeta$, all exist and are finite. From the sufficiency point of view, an arbitrary gauge of $\uppsi$ cannot be verified by the Biot-Savart integral in the absence of $\upomega$ regularity. Let us choose a non-zero constant or any continuous function $f(t)=C=\nabla{\cdot}\uppsi$. Poisson solution (\ref{vp}) becomes a contradiction. 

For $\uppsi$ free of gauge constraint, we again take curl to derive 
\begin{equation} \label{vv}
	\Delta(\nabla{\times}\uppsi)=\Delta\bfu=-\nabla{\times}\upomega,
\end{equation}
which is justifiable for $\upomega \in L^2 \cap C^2$. In square-integrable flow-fields or the Leray-Hopf energy bound ($\bfu, \upomega \in L^2(\real^3)$), the elliptic system (\ref{vpo})-(\ref{vp}) and the $\uppsi{-}\upomega$ identity constitute a pair of illogical relationships. To understand this peculiar result, we draw our attention to the fact that supplementary function $\uppsi$ is not a definitive quantity in physics. Inevitably, the orthodox scheme of evaluating a singular integral for gradient $\nabla\bfu$ necessitates technical inconsistencies or analytical hypotheses about flow physics, see, for instance, \S2.4 of Majda \& Bertozzi (2002) for a pr{\'e}cis of the related issues. 

The significance is that vorticity consists of fluid elements, and has structures. It is exactly due to their propensity for shearing deformation that expedites scale multiplications. From a  heuristic point of view, all small-scale shears participate in moderating the vortex stretching, $(\upomega{\cdot}\nabla)\bfu$, at every instant of the dynamic evolution. Thus, there are no physical reasons that prevent advances of locally subdued vortices in the event of extreme stretching, which counter-balance possible growth in the gradient. In the absence of the singular operator, there are no recipes for flow blow-up. By comparison, theoretical evaluation of the vorticity on $\real^3$ is less troublesome with the velocity-divorticity relation (\ref{vv}). For turbulence, the assumption of homogeneity no longer makes sense.

\vspace{1cm}
\begin{acknowledgements}
\noindent 
26 March 2021

\noindent 
\texttt{f.lam11@yahoo.com}
\end{acknowledgements}

\end{document}